\documentclass[]{aa} 
\usepackage[varg]{txfonts}
\usepackage[colorlinks=true, urlcolor=blue,citecolor=blue,linkcolor=blue]{hyperref}
\usepackage{graphicx}
\usepackage{natbib}
\usepackage{subcaption}
\usepackage{wasysym}
\usepackage{footnote}
\def\kms{\mbox{km s$^{-1}$}}
\def\halpha{{H$\alpha$}}

\def\FeI{\ion{Fe}{I}}
\def\CaIIH{\ion{Ca}{II}~H}
\def\CaII{\ion{Ca}{II}}
\def\CaIII{\ion{Ca}{III}}
\def\CIV{\ion{C}{IV}}
\newcommand{\bea}{\begin{eqnarray}}
\newcommand{\eea}{\end{eqnarray}}
\newcommand{\grass}{}

\begin{document}

\title{The chromosphere above a $\delta$-sunspot in the presence of fan-shaped jets}
\titlerunning{$\delta$-sunspot with fan-shaped jets}
\authorrunning{Robustini, Leenaarts and De la Cruz Rodr\'{i}guez}

\author{Carolina Robustini\inst{1},
 Jorrit Leenaarts\inst{1}, 
 \and
  Jaime de la Cruz Rodr\'{i}guez\inst{1}}

\institute{Institute for Solar Physics, Department of Astronomy,
  Stockholm University,
AlbaNova University Centre, SE-106 91 Stockholm Sweden \email{carolina.robustini@astro.su.se}}

\date{Received; accepted}
\abstract

\abstract {$\delta$-sunspots are known to be favourable locations for fast and energetic events like flares and CMEs. The photosphere of this type of sunspots has been  thoroughly investigated in the past three decades. The atmospheric conditions in the chromosphere are not so well known, however.
}
   {This study is focused on the chromosphere of a $\delta$-sunspot that harbours a series of fan-shaped jets in its penumbra . The aim of this study is to establish the magnetic field topology and the temperature distribution in the presence of jets in the photosphere and the chromosphere.
   }
   {We use data from the Swedish 1-m Solar Telescope (SST) and the Solar Dynamics Observatory. We invert the spectropolarimetric FeI\ 6302~\AA\ and \CaII ~8542~\AA\ data from the SST using the the non-LTE inversion code NICOLE to estimate the magnetic field configuration, temperature and velocity structure in the chromosphere.
   } 
   {A loop-like magnetic structure is observed to emerge in the penumbra of the sunspot. The jets are launched from the loop-like structure.  Magnetic reconnection between this emerging field and the pre-existing vertical field is suggested by hot plasma patches on the interface between the two fields. The height at which the reconnection takes place is located between $\log \tau_{500} = -2$ and $\log \tau_{500} = -3$. The magnetic field vector and the atmospheric temperature maps show a stationary configuration during the whole observation. 
   }
   {} 
 

\maketitle
\section{Introduction}


The chromosphere above sunspots exhibits many dynamic phenomena, such as umbral flashes, running penumbral waves, and various types of jets.

Chromospheric fan-shaped jets launched from sunspots have been reported by several authors
\citep{1973SoPh...32..139R, 2001ApJ...555L..65A, 2009ApJ...696L..66S, 2016A&A...589L...7H, 2016A&A...590A..57R, 2016ApJ...826..217L, 2016ApJ...833L..18Y}. 
From these previous observations, we know that the length of these jets is of tens of Mm. They have an average velocity of 100-200~\kms\ and can last for more than one hour. They exhibit a sideways motion that has been observed 
 \citep{1992PASJ...44L.173S,2007PASJ...59S.771S} 
 and simulated 
  \citep{2008ApJ...673L.211M,2013ApJ...771...20M} 
 also in anemone jets. The jets appear dark in \halpha\ and  may exhibit a bright front in the EUV lines. In 
 \citet{2009ApJ...696L..66S} 
 and
  \citet{2016A&A...590A..57R}
 the jet footpoints appear bright in \CaIIH\ and \halpha\  respectively. It has been suggested that the driver of this type of jets is magnetic reconnection and that, consequently, the bright footpoints are the result of local plasma heating. 

\citet{2011ApJ...726L..16J} 
have reproduced the structure of a fan-shaped jet in a 3D simulation of magnetic reconnection. The fan structure that they simulate is caused by the sheared guide field lines that thread through the current sheet, and the jets are accelerated first by the magnetic pressure gradient and then by gas pressure gradients.


These jets are recurrently launched above sunspot structures. The majority
\citep{1973SoPh...32..139R, 2001ApJ...555L..65A, 2009ApJ...696L..66S, 2016A&A...590A..57R, 2016ApJ...833L..18Y} 
have been reported on light-bridges. There are nonetheless some exceptions. In one of the  observations of 
\citet{1973SoPh...32..139R} 
(MW 18594) the jets are rooted in the penumbra of a negative sunspot group harbouring a positive polarity patch that weakens when jets start appearing. In the observations of
 \citet{2016A&A...589L...7H} 
 the jets are launched from an apparent positive polarity field region between two distinct regular $\alpha$-negative sunspots. 

In this paper we report on fan-shaped jets observed in the penumbra of a $\delta$-sunspot. This kind of sunspot configuration consists of umbrae of both polarities sharing the same penumbra \citep{1960AN....285..271K}. $\delta$-sunspots can harbour strong current densities
\citep{2003A&ARv..11..153S}
 and are often associated with flaring activity
 \citep{1987SoPh..113..267Z,2000ApJ...540..583S}. 
It has been suggested that the complex topology of $\delta$-sunspots originates from the emergence of twisted flux tubes 
\citep{1991SoPh..136..133T,2002ApJ...572..598K}.
\citet{2015ApJ...813..112T} 
simulated the formation of a $\delta$-sunspot configuration from the emergence of an unstable kinked flux tube which spontaneously develops into a quadrupole. The four polarities do not appear all together. First to appear is a main pair. The arcade connecting the first pair expands and plasma accumulates on its top. Eventually this leads to the submergence of the magnetic field and the appearance of a second pair of polarities between the main pair. The emerging flux tube employed by
 \citet{2015ApJ...813..112T} 
had a single buoyant segment. 

In addition, $\delta$-sunspots have been simulated using two buoyant segments in a twisted flux emerging at the same time 
\citep{2015ApJ...806...79F}.
The submergence of the magnetic field predicted by these two model could find its observational proof in the downflow observed at the PIL by  
\citet{1994ApJ...425L.113M}.
\citet{2014ASPC..489...39B}  
and 
\citet{2016ApJ...818...81J} 
employed near-infrared spectropolarimetry to retrieve the magnetic and dynamical properties of  $\delta$-sunspots in the photosphere. 
\citet{2014ASPC..489...39B} 
report on the presence of an upflow aligned with the polarity inversion line (PIL) of the spot and some photospheric brightenings. 
\citet{2016ApJ...818...81J}
found an intensification of the transverse magnetic field at the PIL. A similar intensification can also be found in 
\citet{2014ApJ...789..162C}.
While the magnetic topology of $\delta$-sunspots has been largely studied in the photosphere, we know very little about its configuration in the chromosphere.
In this paper we will present the results of a study of the chromosphere above a $\delta$-sunspot in the presence of fan-shaped jets, using polarimetric data inversion.



\section{Observations and data reduction}
The target of the observations is a $\delta$-sunspot located in the active region NOAA 11791, observed on 2013 July 15 from 07:18 to 08:24 UT.
The coordinates in the middle of the time series are $15.26^{\circ}~\textrm{S},~16.69^{\circ}~\textrm{E}$, with an observing
angle of $27^\circ$ ($\mu = 0.89$).

Our observations were carried out at the Swedish 1-m Solar Telescope 
\citep[SST,][]{2003SPIE.4853..341S} 
using the CRisp Imaging SpectroPolarimeter 
\citep[CRISP,][]{2008ApJ...689L..69S} 
along three different line profiles:
\begin{itemize}
\item[•]\halpha\ 6563~\AA, at 13 positions between 6561.45 and 6564.55~\AA,
\item[•]\FeI\ 6301-6302~\AA, at 18 positions between 6300.45 and 6302.10~\AA,
\item[•]\CaII\ 8542~\AA, at 21 positions between 8540.25 and 8543.75~\AA.
\end{itemize} 
The cadence between two complete profile scans along the same profile is 27~s. The pixel size and the spectral resolution at 630~nm are 0\farcs059 and $\mathrm{R}=\lambda / \delta\lambda \approx114000$ respectively. 

We recorded full Stokes vector data for \FeI\ 6301-6302~\AA\ and \CaII ~8542~\AA. These lines, with effective Land\'e factor of 2.5 and 1.1 respectively, are good diagnostic tools for the magnetic field in the photosphere (\FeI) and chromosphere (\CaII). The CRISP data reduction followed the pipeline described in
\citet{2015A&A...573A..40D}
which includes image restoration with Multi-Object Multi-Frame Blind Deconvolution
\citep[MOMFBD,][]{2005SoPh..228..191V}.
In order to avoid degradation of the signal-to-noise ratio from interpolation noise, the three datasets were spatially aligned using as reference cube the \CaII~8542~\AA\ data which has a weaker polarimetric signal compared to \FeI\ 6301-6302~\AA.

We also made use of co-observations of the Helioseismic and Magnetic Imager 
\citep[HMI,][]{2012SoPh..275..207S} 
and the Atmospheric Imaging Assembly 
\citep[AIA,][]{2012SoPh..275...17L}
on board of the Solar \grass{Dynamics} Observatory 
\citep[SDO,][]{2012SoPh..275....3P}. 

AIA data (with 12~s cadence) and HMI data (with 48~s cadence) have been  aligned and resampled in space and time to match the cadence and pixel size of the SST data using the routines developed by R. J. Rutten\footnote{\url{http://www.staff.science.uu.nl/~rutte101/rridl/sdolib/}}.  
The alignment has been done by first rotating the SDO subfield of interest to the same orientation as the SST data. Then a cross correlation between the SST \halpha\ wide-band and the HMI continuum data is performed. The accuracy of the co-alignment is on the order of an SDO pixel size ($0\farcs5$).

\begin{figure}
\includegraphics[scale=1]{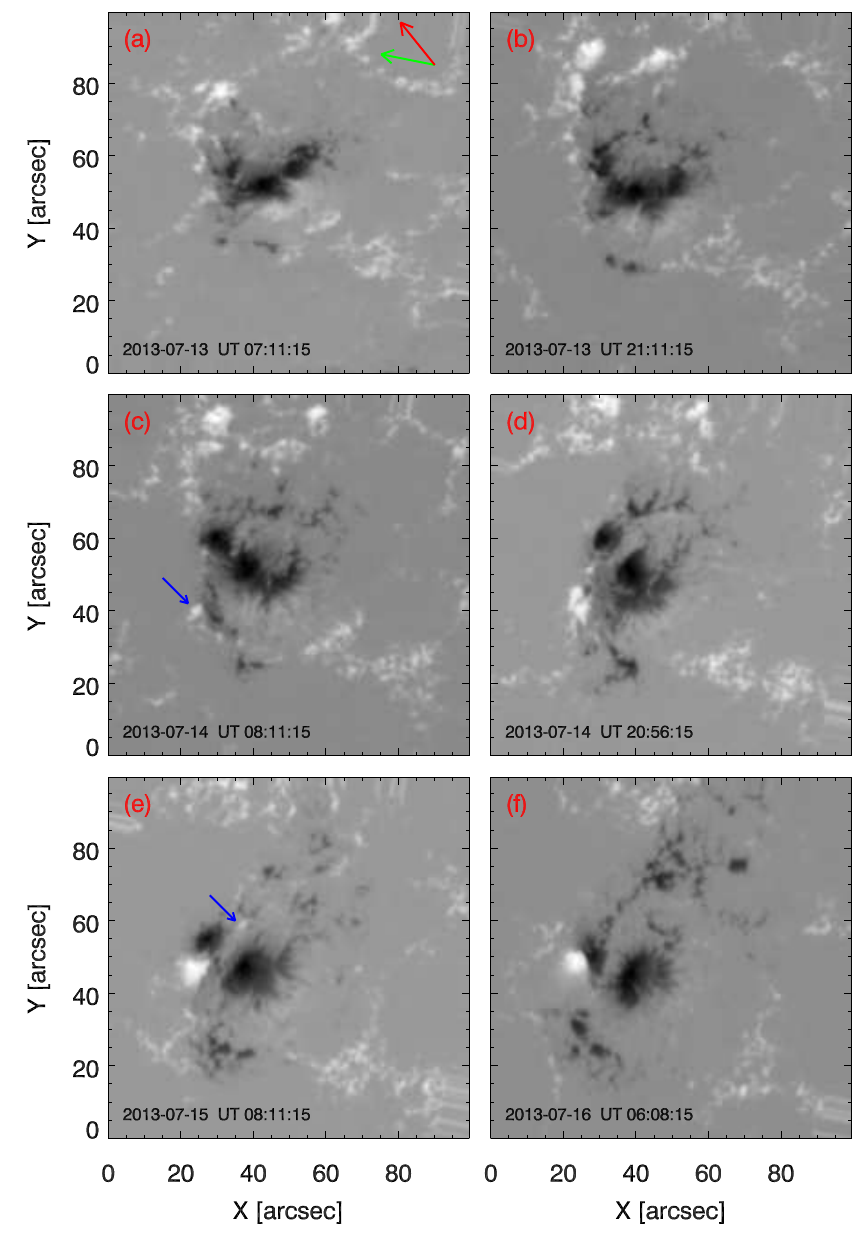}\caption{The time evolution of the LOS magnetic field (SDO/HMI) of a sunspot group located in the active region NOAA 11791. Positive flux emergence is highlighted by the blue arrows in panels~(c) and~(e). The green and the red arrows point towards the \grass{north} and the disk centre respectively.}\label{fig:hmievol}
\end{figure}

Figure~\ref{fig:hmievol} shows the time evolution of the line of sight (LOS) magnetic field provided by HMI. Initally, the whole sunspot group has negative polarity (panels a--b). Positive flux emergence appears on 2013 July 14 (arrow in panel~c) and forms a small sunspot of opposite polarity in the penumbra of the main sunspot. A second smaller area of positive flux emergence appears at the same time of the observations (arrow in panel~e). This positive flux cancels with the close negative polarity, leaving no trace the day after (panel~f). 

\begin{figure}
\includegraphics[scale=1]{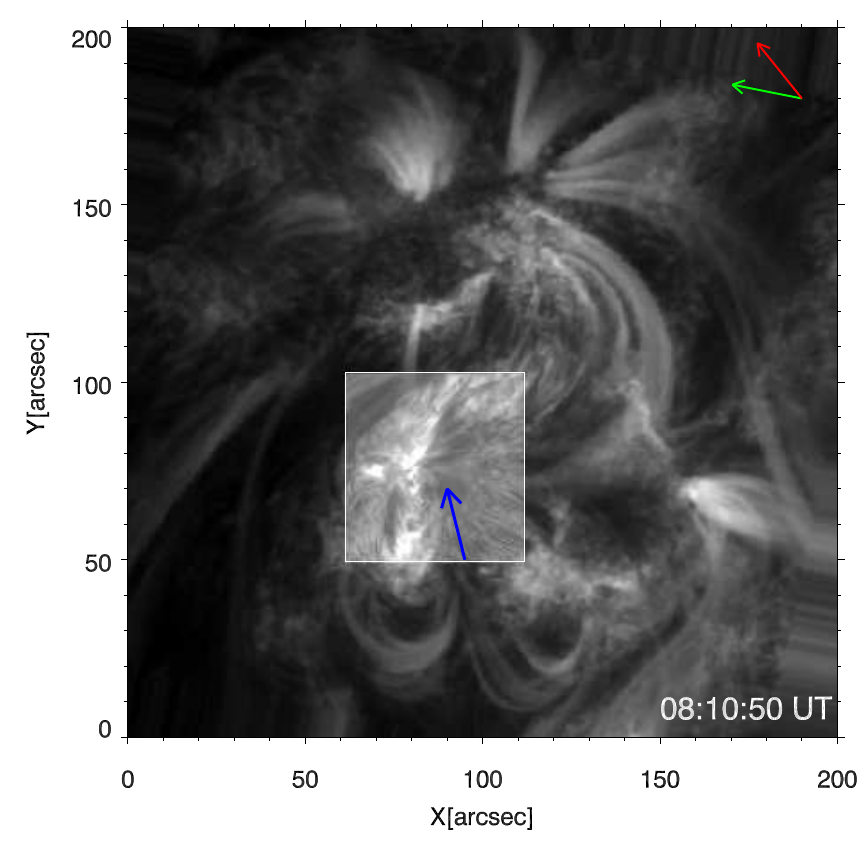}\caption{AIA~171 image of the active region (AR) NOAA 11791. The square inset indicates the FOV of the SST observations (see Figure~\ref{fig:overview}), and shows the \halpha\ line core at 08:10:50~UT. The blue arrow highlights the location of the fan-shaped jets. The green arrow points towards the \grass{north} and the red arrow towards the disk centre.}\label{fig:loop}
\end{figure}

Figure~\ref{fig:loop} shows the AR of interest in the AIA~171 channel. A coronal loop connects the positive polarity of the AR with the negative one, where the CRISP field of view (FOV) is located. The CRISP FOV is indicated by a white box that contains a superposition of AIA~171 and \halpha\ line core images. The blue arrow highlights the position of the fan-shaped jets featured in our observations and it shows that the entire jet structure is aligned with the direction of the coronal loop.

\section{Results}

\begin{figure*}
\includegraphics[scale=1]{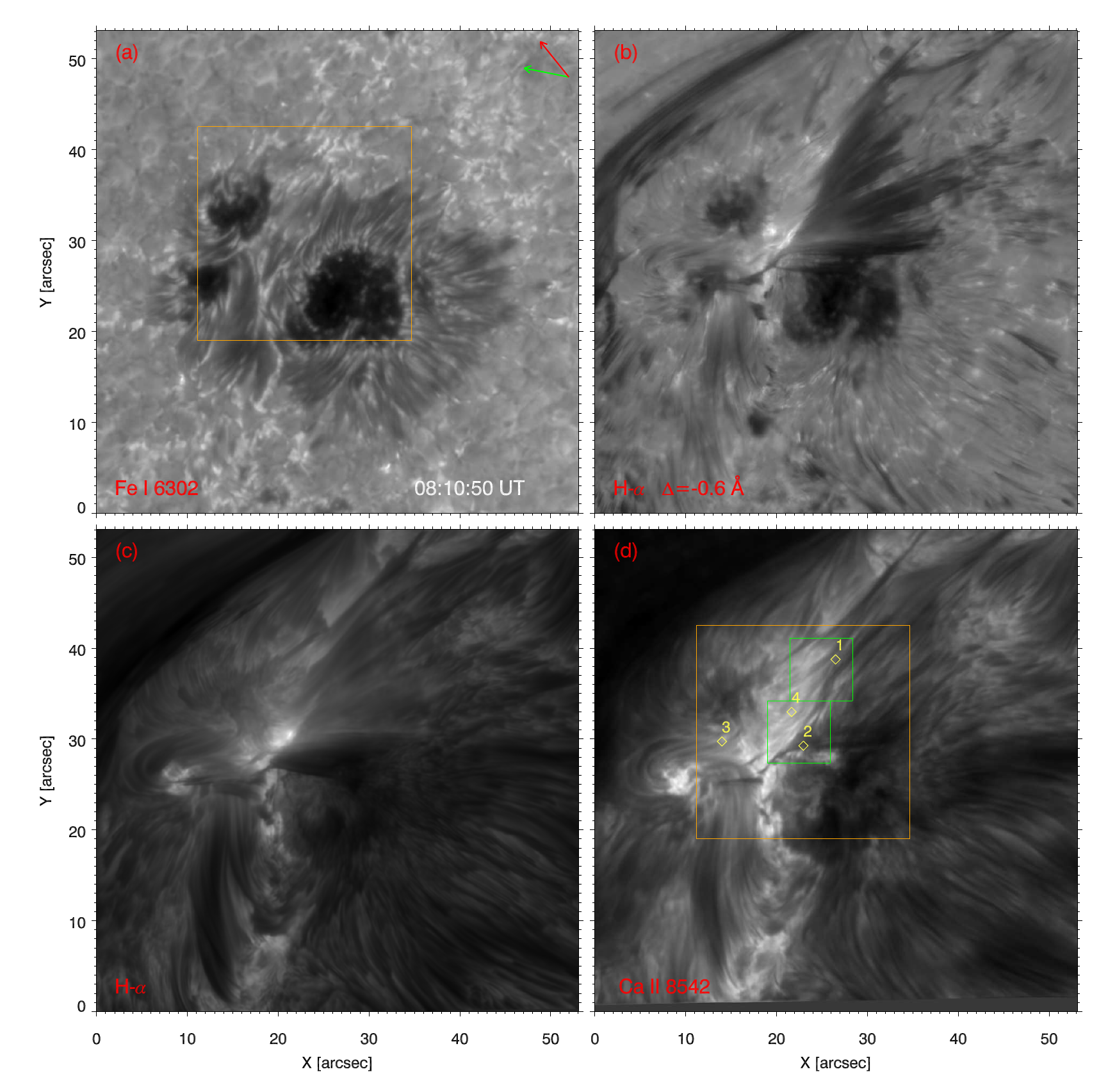}\caption{Sample of the dataset at 08:10:50~UT. (a) Line core of \FeI\ 6302~\AA , (b) \halpha\ blue wing at $\Delta \lambda =$ -0.6~\AA , (c) \halpha\ line core and (d) \CaII\ ~8542~\AA\ line core. The orange box is the subfield chosen for the inversion of this time step. The smaller green boxes are the subfields for which the entire time series has been inverted. The numbered diamond symbols indicate the location of the selected Stokes profiles shown in Figure~\ref{fig:stokes}. The green arrow points towards the \grass{north} and the red one towards the disk centre. \grass{The temporal evolution can be shown in a movie available online. In the middle of the time series, the field of view is shifted.}}\label{fig:overview}
\end{figure*}

Figure~\ref{fig:overview} displays a sample of the dataset at 08:10:50~UT.  Panel~(a) shows the photosphere observed in the line core of \FeI\ 6302~\AA . The penumbra of the main sunspot appears slightly twisted counter-clockwise, as do the smaller spots. \FeI\ 6302~\AA\ does not exhibit any remarkable activity. On the contrary, dark recurrent fan-shaped jets and bright footpoints appear in the penumbra between the main sunspot and the other spots in the \halpha\ line core (c) and wings (b). The jets appear less visible in the core of \CaII ~8542~\AA . Only the fronts are dark and have good contrast while the rest of the jets have line cores in emission. The plasma is mainly ejected in the S-W direction except for an apparently shorter series of jets located at X=16$\arcsec$,Y=24$\arcsec$ (panels c and d). 
In the associated animation it is possible to observe a transversal motion of the jets along the bright lane of the footpoints.
During the entire observation, the plasma ejections are accompanied by a flare-like brightening located on the left side of the positive sunspot, as visible in the chromospheric images in panel c and d.

\begin{figure}
\includegraphics[scale=1]{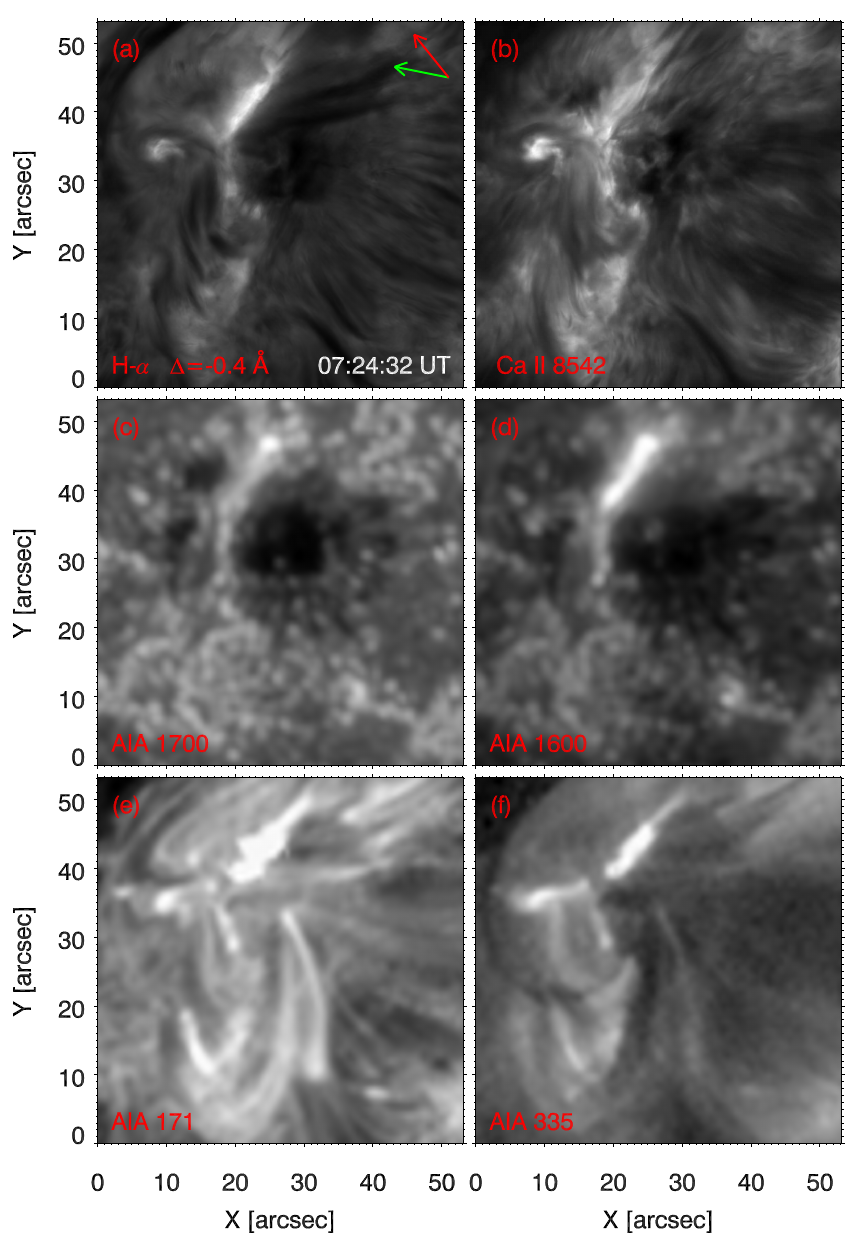}\caption{Bright event at the jet footpoint at 07:24:32~UT. Panel (a) shows the blue wing of \halpha\ at $\Delta \lambda =$ -0.4~\AA , panel (b) displays the line core of \CaII\ ~8542~\AA. Panel(c) and (d) show the logarithm of the intensity in the AIA 1700~\AA\ and 1600~\AA\ channels respectively. Panel (e) and (f) show the AIA EUV channels 171~\AA\ and 335~\AA . Panels (c--f) are spatially and temporally co-aligned to CRISP dataset. The green arrow points towards the \grass
{north} and the red arrow towards the disk centre. \grass{The temporal evolution can be shown in a movie available online.}}\label{fig:allchannels}
\end{figure}

\begin{figure}
\includegraphics[scale=1]{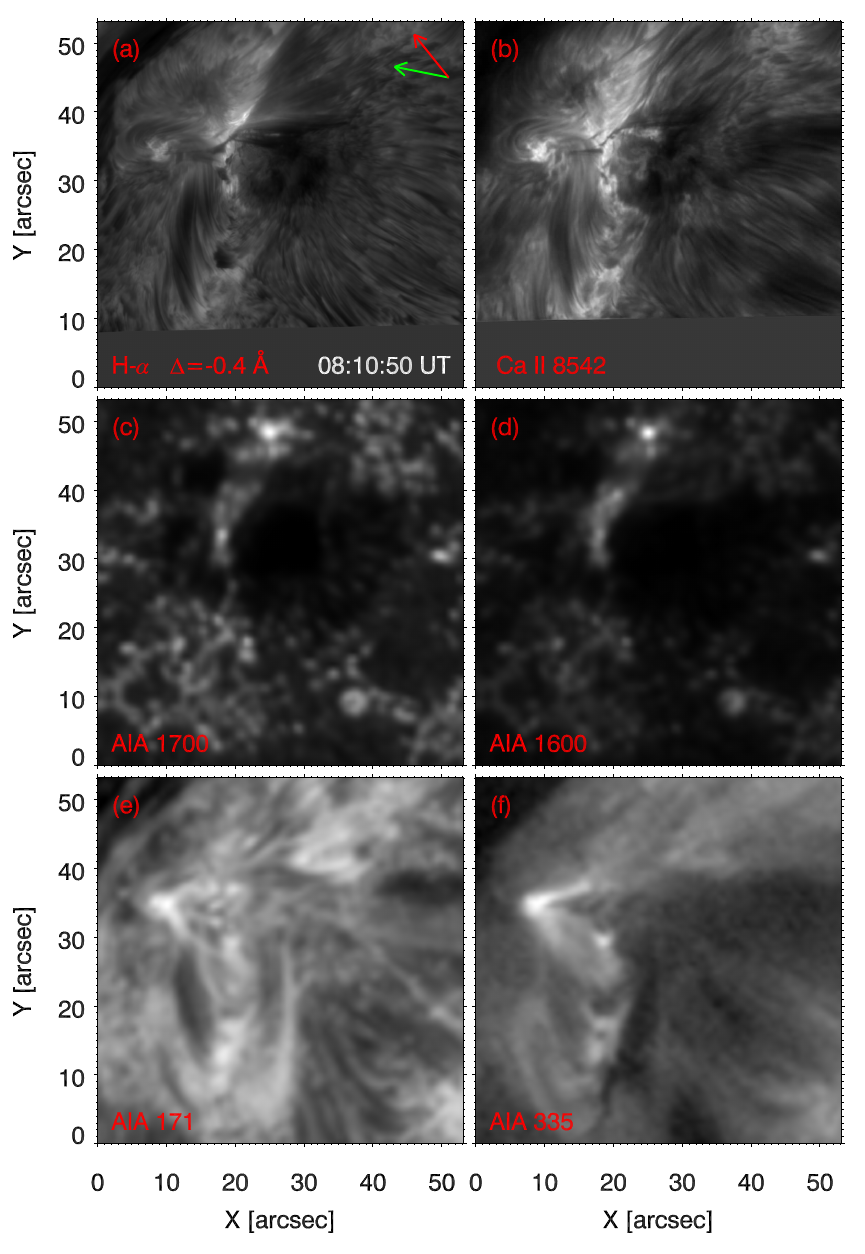}\caption{Same as Figure~\ref{fig:allchannels} but at 08:10:50~UT and (c) and 
(d)  have a linear brightness scale instead of a logarithmic one. \grass{The temporal evolution can be shown in a movie available online.}}\label{fig:allchannels_114}
\end{figure}

Figure~\ref{fig:allchannels} shows SST and AIA images of one of the brightest events of the entire time series. The bright event is located at the footpoint of the jets as clearly visible in \halpha\ (a). \grass{Unlike the constant brightening on the left of the sunspot group, the footpoint intensity is highly variable with time.}

The brightening has the same spatial extent  in the 1600~\AA\ channel (d) while in 1700~\AA\  (c) only one smaller round brightening appears at the top of the bright lane. The 1700~\AA\ opacity is dominated by the \ion{Si}{I} continuum and a multitude of UV lines, and the intensity forms in the photosphere and low chromosphere
\citep{2005ApJ...625..556F}.
It is not sensitive to transition region temperatures. The AIA 1600~\AA\, band is sensitive both to photospheric and transition region temperatures owing to the \CIV\ lines  located around 1550~\AA. This suggests that the plasma at the base of the jets can have a transition-region like temperature, much higher than normally found in the chromosphere. \CaII ~8542~\AA\ in panel~(b), although having similar appearance to \halpha, does not show such a strong brightness. This can be explained by the higher opacity of \halpha\ compared to \CaII ~8542~\AA\ at temperatures well above 20~kK
\citep[e.g.][]{2012A&A...539A..39C,2016A&A...590A.124R,2017A&A...598A..89R}.
The same extreme event is visible in the EUV channels of AIA 171~\AA\ (e) and 335~\AA\ (f). In 171~\AA\ the jet fronts are brighter than the jet bulk, as already seen in previous observations \citep{2016A&A...590A..57R}. The jets and the front are darker in 335~\AA\, for which the characteristic temperature is $2.5 \times 10^6$~K \citep{2012SoPh..275...17L}, that is one order of magnitude larger than 171~\AA. This poses an upper boundary to the temperature of the jet fronts.

The panels of Figure~\ref{fig:allchannels_114} show the same wavelengths of Figure~\ref{fig:allchannels} at 08:10:50 UT (same time of Figure~\ref{fig:overview}). 
For this time-step, the \halpha\ footpoint brightening is no longer visible in the EUV channels (e-f). 1600~\AA\ has a quite similar appearance to 1700~\AA\ except for a small brightening at X=21$\arcsec$, Y=39$\arcsec$ that corresponds to the core of the \halpha\ brightening. This feature is better displayed in Figure~\ref{fig:tempvssdo_111_114}-c.  

Unlike the extreme event of Figure~\ref{fig:allchannels}, the time-step shown in Figure~\ref{fig:allchannels_114} is representative of the entire time series. The jet footpoints typically show similar scenes in AIA 1600~\AA\ and 1700~\AA. The comparison between Figure~\ref{fig:allchannels} and \ref{fig:allchannels_114} shows that there is a preferred path along which heating, and consequent brightenings, develop and that there are locations that are constantly active along this path.  

The different appearance of the fine structure, visible in panels a and b in both figures, is due to the variable seeing conditions in the Earth's atmosphere that change the quality of the data despite the image restoration.

\subsection{Data inversion}
To reconstruct the structure of the atmosphere we have used the non-LTE inversion code NICOLE \citep{2015A&A...577A...7S}. 
We have included the effect of \ion{Ca}{ii} isotopic splitting in our calculations, which introduces a red-wing asymmetry in the 8542 line profile \citep{2014ApJ...784L..17L}. The transfer equation is solved using a cubic DELO-Bezier solver \citep{2013ApJ...764...33D} and a regular depth-scale grid of 5 points per decade, which ensures a sufficiently accurate solution when computing the four Stokes parameters \citep{2017ApJ...840..107J}. For further details we refer to the code description paper.

The field of view that we have chosen is indicated by the orange boxes in Figure~\ref{fig:overview}-(a) and (d). The inversion of such a large field of view (FOV) is computationally demanding. Therefore we selected only one time step, at 08:10:50~UT (see Figure~\ref{fig:overview}) that exhibits good seeing in all the wavelength positions for both \FeI\ 6301-6302~\AA\ and \CaII ~8542~\AA. 
In addition we inverted the \CaII ~8542~\AA\ data for the entire time series for two smaller FOVs indicated in by the green boxes Figure~\ref{fig:overview}-(d).

The starting guess model for all the inversions is a FALC atmosphere \citep{1993ApJ...406..319F} with enhanced gas pressure at the upper boundary that accounts for the typical values of active regions. 
For the inversion of \FeI\ 6301-6302~\AA\ we have initialised the three components of the magnetic field ($B_x,B_y,B_z$) with a constant value of 500 G. We have used 4 equidistant nodes in temperature, 2 in velocity, 2 in $B_x,B_y,B_z$ and 1 in microturbulence. For the inversion of both the smaller and larger FOVs of \CaII ~8542~\AA\ we have used the same FALC atmosphere and a first estimate of the magnetic field obtained by the weak field approximation. The temperature has been fitted with 5 non-equidistant nodes, with $\log \tau_{500}$ at -7.0, -4.2, -2.9, -1.5, and -0.2, that have been empirically selected to obtain the best fit of the Stokes parameters. We have used 2 equidistant nodes for the velocity and the microturbulence. A good fit of the parameters can be obtained with just two equidistant nodes in $B_x,B_y,B_z$. However we decided to apply 3 equidistant nodes to obtain magnetic field maps that look smoother. 

The inversion of polarimetric data in \CaII ~8542~\AA\ can be very challenging compared with photospheric inversions in the \FeI\ 6301-6302~\AA lines. The main reason is the lower signal-to-noise ratio of chromospheric observations because of the weaker magnetic field. 
The reliability of the results is mainly guaranteed by the goodness of fit between the observed and the synthetic Stokes parameters retrieved from inversion. However, also the physical meaning of the quantities characterising the inferred model atmosphere have to be considered 
in order to establish whether the results can be accepted or not.

\begin{figure}
\includegraphics[scale=1]{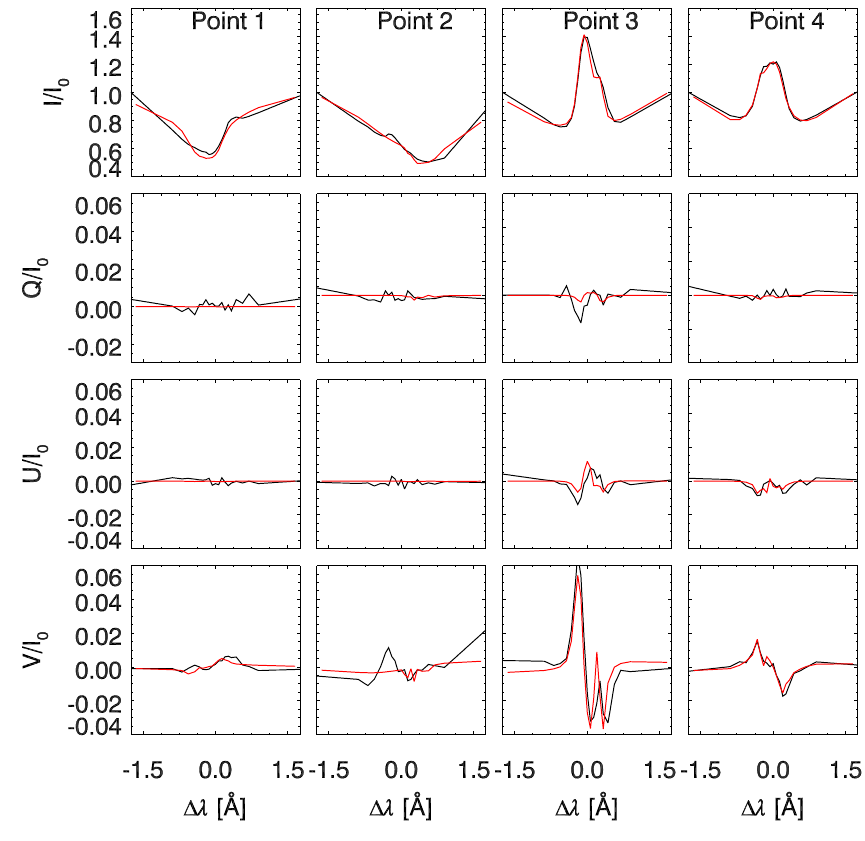}\caption{Examples of typical Stokes profiles in the \CaII ~8542~\AA line. The locations of the profiles are indicated in Figure~\ref{fig:overview}d and~\ref{fig:magnetograms_111_114}. The black and the red curves represent the observed profiles and their fit respectively. A description of the characteristics of each point is given in the text. }\label{fig:stokes}
\end{figure}

Figure~\ref{fig:stokes} shows four examples of Stokes parameter profiles (black curve) and the fit (red curve) of the \CaII ~8542~\AA\ dataset. These profiles represent the time average of four consecutive frames ($27\times 4$~s). This averaging increases the signal-to-noise ratio but on the other hand can hide fast evolution of the atmosphere.
The locations of these examples are indicated in Figure~\ref{fig:overview}-(d). We have selected them because they show typical profiles of the region in which they are located.  

Point 1 is located in the dark part of the jet. The intensity is blue-shifted and the circular polarisation $V$ has several lobes, consistent with a region of upwards and downwards plasma motions coexisting in the same resolution element and inhomogeneities in the velocity. 
Point 2 represents a dark region of the jet too, but its profile is red-shifted. The circular polarisation $V$ has a clear asymmetry that NICOLE cannot fit properly. Both point 1 and 2 have a small intensity bump in the red and in the blue wing respectively at the same wavelength of the maximum of the circular polarisation.
Point 3 has been selected to compare its strong signal with the other three. It is located in the twisted penumbra of the negative small sunspot. The intensity profile is reversed in the core giving rise to an opposite sign of the stokes $V$ signal
  \citep{1997A&A...324..763S}. 

The linear polarisation ($Q$ and $U$) signal is above the typical noise level $10^{-3}$ of an imaging spectropolarimeter \citep{2015SSRv..tmp..115L}. Stokes $V$ exhibits asymmetry as in the case of the accelerated plasma of points 1 and 2, which appear clearer thanks to a stronger magnetic field. 
Point 4 shows the Stokes profiles of the bright part of the jets, close to the jet footpoints. The profiles of this region differs significantly from points 1 and 2: the signal in \grass{$U$} is higher with respect to the dark part of the jets and Stokes $V$ is quite symmetric and has a third central lobe.
The former can be a sign of inhomogeneities in the magnetic field
 \citep{2015AdSpR..56.2305L} 
and is consistent with magneto-optical effects. These effects becomes more important as the magnetic field gets stronger and more inclined 
\citep{1982SoPh...78..355L}, 
which is the case for point 4 (see the magnetograms in Figure~\ref{fig:magnetograms_111_114}).

\subsubsection{Magnetic field}

The magnetic field values that the inversion returns are affected by Zeeman azimuth ambiguity in the transversal component. There are several possibilities to get rid of this ambiguity \citep{2006SoPh..237..267M}. We have chosen the minimum energy method (MEM) proposed by \citet{1994SoPh..155..235M}, using the implementation of \citet{2014ascl.soft04007L}. This method has been demonstrated to be a promising solution for the photosphere of complicated active regions \citep{2006SoPh..237..267M}. 
We have retrieved the azimuth map of the \FeI\ 6301-6302~\AA\ at $\log \tau_{500} = -1$ , that is the depth point at which the line is most sensitive. The noise in the chromospheric magnetic field of \CaII ~8542~\AA\ prevents the use of the MEM routine. Thus, we have applied the principle of the acute angle method \citep{1985tphr.conf..313S}, comparing the direction of the photospheric field with the chromospheric field at  $\log \tau_{500} = -3$. We have used the same approach to retrieve the azimuthal direction at all the other depth points, both in \FeI\ 6301-6302~\AA\ and \CaII ~8542~\AA. 

\begin{figure*}
\includegraphics[scale=1]{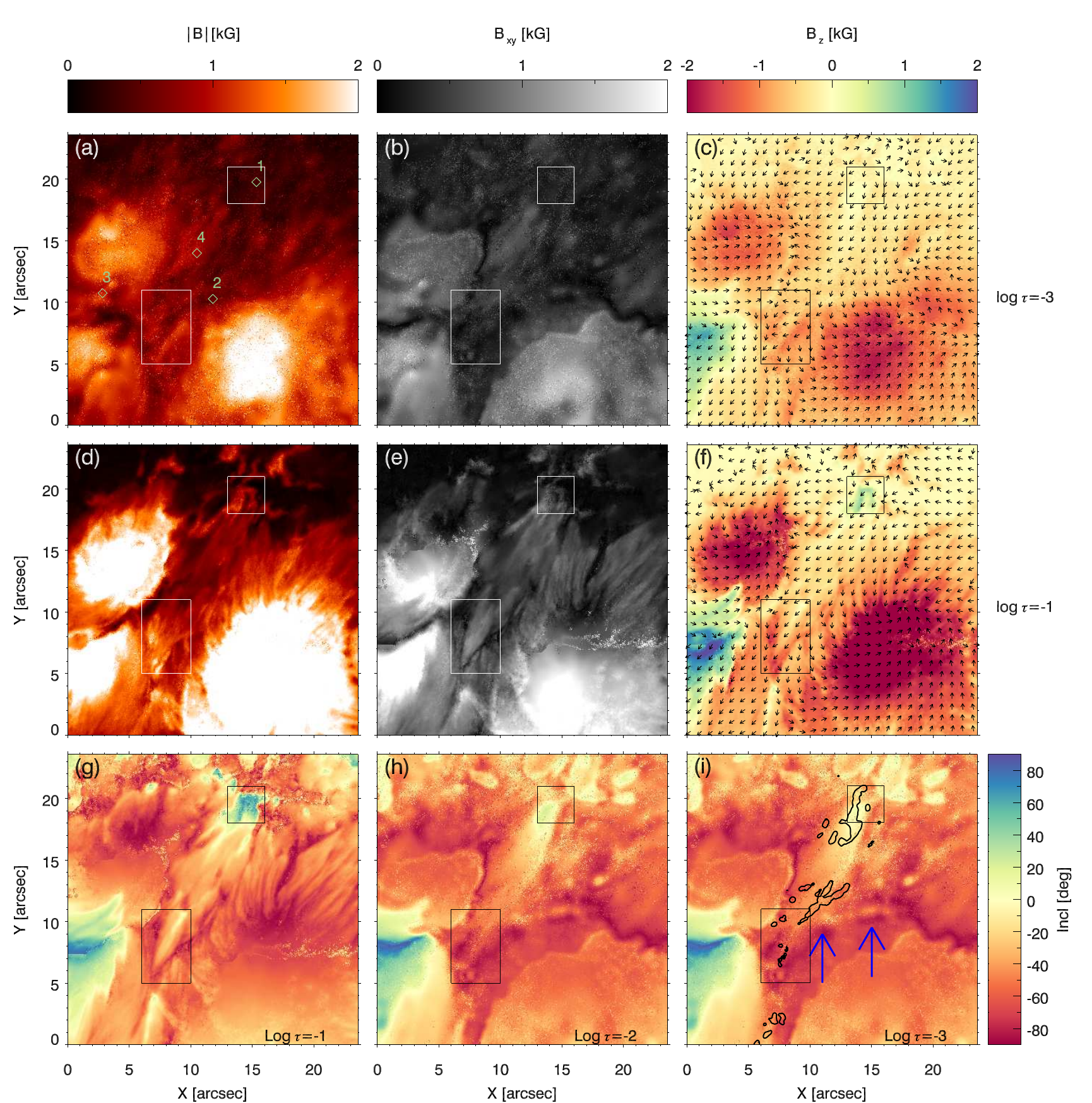}\caption{\textit{Top row:} the total (a), transversal (b) and longitudinal (c) magnetic field in the chromosphere at $\log \tau_{500} = -3$. 
\textit{Middle row:} the total (d), transversal (e) and longitudinal (f) magnetic field in the photosphere at $\log \tau_{500} = -1$.
\textit{Bottom row:} inclination of the magnetic field at $\log \tau_{500} = -1$ (g),$\log \tau_{500} = -2$ (h) and $\log \tau_{500} = -3$ (i). 0 deg means horizontal magnetic field vector. The inclination in panel (g) is derived from the inversion of \FeI\ 6302~\AA\ while panels (h) and (i) are obtained from \CaII ~8542~\AA\ inversions. The black contour in panel (i) indicates temperatures larger than 8~kK at $\log \tau_{500} = -3$. The blue arrows in panel i highlight the imprint of part of the jets. The two rectangles appearing in all the panels show the locations of the footpoints of the emerging loop structure. The numbers in panel a indicate the location of the Stokes profiles shown in Figure~\ref{fig:stokes}.}\label{fig:magnetograms_111_114}
\end{figure*}

Figure~\ref{fig:magnetograms_111_114} displays the  magnetic field maps corrected for azimuth ambiguity and rotated to the local reference system, where the z-axis corresponds to the local solar vertical. The \grass{first and the second row} of Figure~\ref{fig:magnetograms_111_114} show the photospheric and the chromospheric map respectively. The azimuth de-ambiguation failed in the lower-right corner of the FOV that corresponds to the inner part of the main sunspot umbra. The imprint of the azimuth is therefore well visible in both the horizontal field maps (Figure~\ref{fig:magnetograms_111_114}-b and e). This failure, however, occurs in a portion of the FOV in which we are not interested in and it does not affect the rest of the azimuth map. 
The magnetic field maps that have been retrieved are consistent with the scenario in which the magnetic field in the sunspot expands and weakens with height. The chromospheric maps look smoother and have lower field strengths; notice that the three components of the magnetic field have been saturated to 2~kG (-2~kG to +2~kG for the vertical field). These results assure that the field maps of Figure~\ref{fig:magnetograms_111_114} are reliable at large scales. 

The retrieved topology is in agreement with the LOS magnetogram of Figure~\ref{fig:hmievol}-e: it shows a negative polarity main sunspot and a pair of positive and negative smaller spots. 
The horizontal field (panels b and e) in between the umbrae is rather strong and it is, in most of the pixels, aligned with the penumbral filaments of the three sunspots. 
The photospheric vertical magnetic field maps (f) shows two strong-field opposite-polarity patches that have been highlighted in Figure~\ref{fig:magnetograms_111_114} by two rectangles. 
According to the LOS field evolution of Figure~\ref{fig:hmievol}, these patches are newly emerged magnetic field concentrations. If we move higher into the chromosphere (c), these two opposite polarities weaken, especially the upper one located outside of the penumbra. 

The inclination of the field is shown in panels g-i. The inclination at $\log \tau_{500} = -1$ (g) is computed from the inversion of \FeI\ 6301-6302~\AA\, while panels h and i are derived from \CaII ~8542~\AA. Panel g and h show a certain continuity that is not obvious for maps retrieved from two different lines. 
The inclination maps shows that the field between the two vertical field patches is rather horizontal, suggesting the presence of a loop structure that flattens moving higher and has its footpoints in the photosphere ($\log \tau_{500} = -1$).
The horizontal field of this emerging loop is surrounded by rather vertical and strong (0.3 vs 1.2 kG) negative field. 
In such a buoyant structure as this loop, one would expect a torque in the magnetic field lines. However, if the field lines of the loop were twisted we would observe opposite polarities in the loop structure, which instead are missing from the maps of Figure~\ref{fig:magnetograms_111_114}-c and e. 
The connectivity of the loop structure differs from the surrounding field, suggesting a different origin. The interface between the two field, where the inclination gradient becomes larger, roughly coincides with the location of temperature enhancement.  Figure~\ref{fig:magnetograms_111_114}-i shows the contour of the regions where the temperature at $\log \tau_{500} = -3$ is larger than 8~kK: the contour patches are closer to the loop base and stretched along the loop. Their position and shape are  stable during the entire time series. 

The structure of the jets is almost completely absent from the magnetic field maps as their imprint on the Stokes parameters is weak. There are some cases, however, where we can recognise the imprint of the jets in the vertical component of the magnetic field and in the magnetic field inclination map (blue arrows in Figure~\ref{fig:magnetograms_111_114}). The location of the jets is characterised by a rather horizontal and weak field. Such an imprint is visible only for those jets in which the \CaII ~8542~\AA\ intensity core is in absorption. The typical Stokes profiles along the jet footprint are represented by the point two of Figure~\ref{fig:stokes}, where the inversion code systematically fails to fit the circular polarization. It is therefore possible that the values of the field retrieved along that part of the jet are actually unrealistic and that the only reason why we have the imprint is because the misfit produces a very weak field superimposed on a strong field region. 

\begin{figure}
\includegraphics[scale=1]{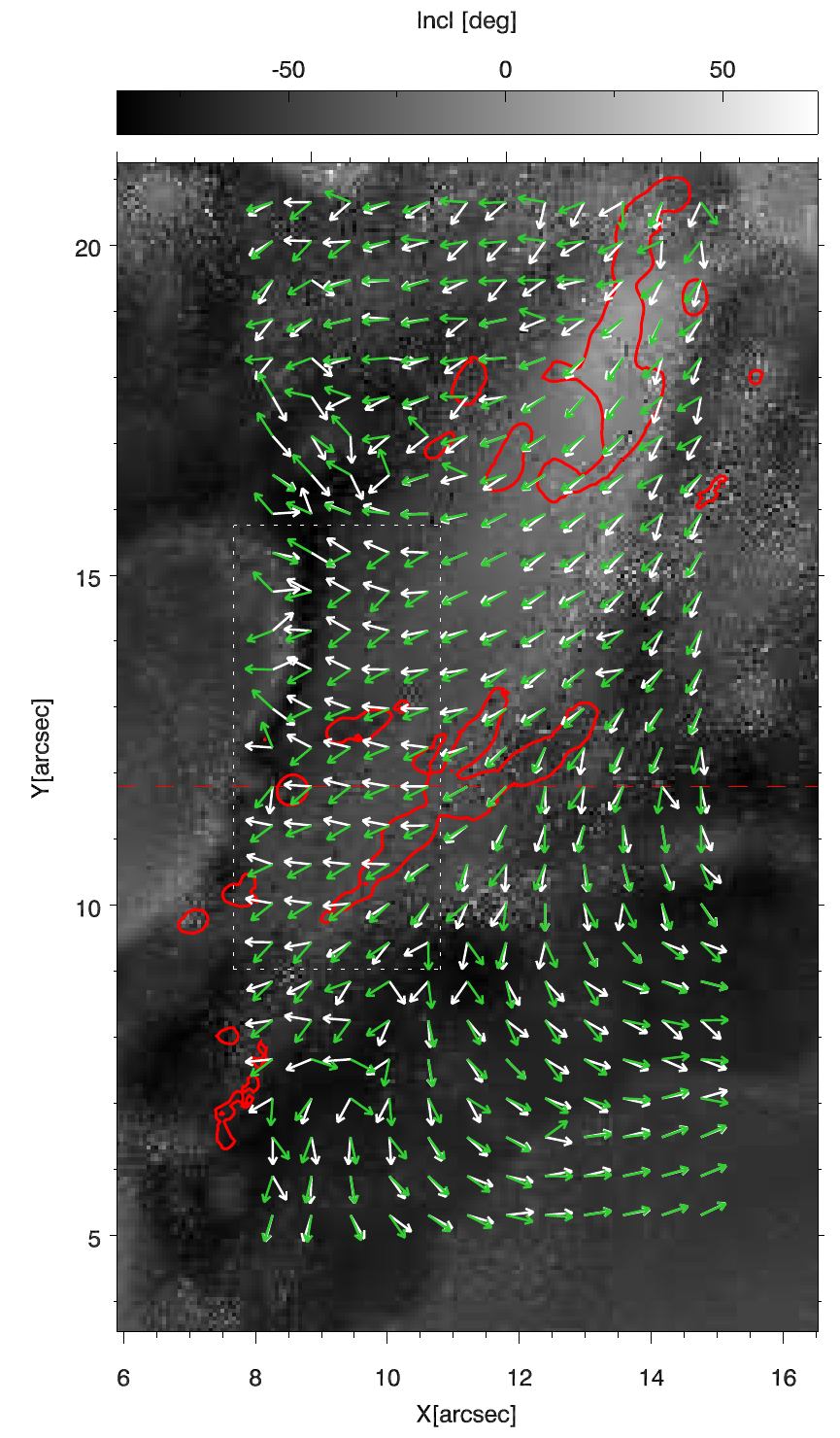}\caption{Zoom of the inclination map of Figure~\ref{fig:magnetograms_111_114}-i. the red contour shows the same temperature contour as in Figure~\ref{fig:magnetograms_111_114}-i. The white arrows show the direction of the horizontal magnetic field in the chromosphere and the green arrows in the photosphere. The dashed rectangle highlights the region where there is a strong change of the horizontal field direction with height. The red dashed line is the vertical cut shown in the upper panels of Figure~\ref{fig:temp_p_cut_111_114}.}\label{fig:shear}
\end{figure}

Figure~\ref{fig:shear} is a zoom of the region including the loop-structure footpoints. It shows the inclination map at $\log \tau_{500} = -3$ and superimposed on it the direction of the horizontal field in the chromosphere (white arrows) and photosphere (green arrows). 
Along the loop, the direction of the horizontal field seems to be quite constant with height since the green and the white arrows are, most of the time, overlapping. However, in the region delimited by the dashed rectangle, the chromospheric horizontal field bends towards the small sunspot and the field direction is no longer constant in height. This indicates that there is a shear along the vertical direction, in the region between the heating patches. Close to the umbra (left side of the dashed rectangle) the direction of the chromospheric horizontal field is visibly incoherent because of the noisy inversion results produced by stray-light on the umbral edges. This example shows how the results of a locally bad inversion can be easily spotted when considered in a larger and reliable context.   

\subsubsection{Temperature and pressure}
\begin{figure}
\includegraphics[scale=1]{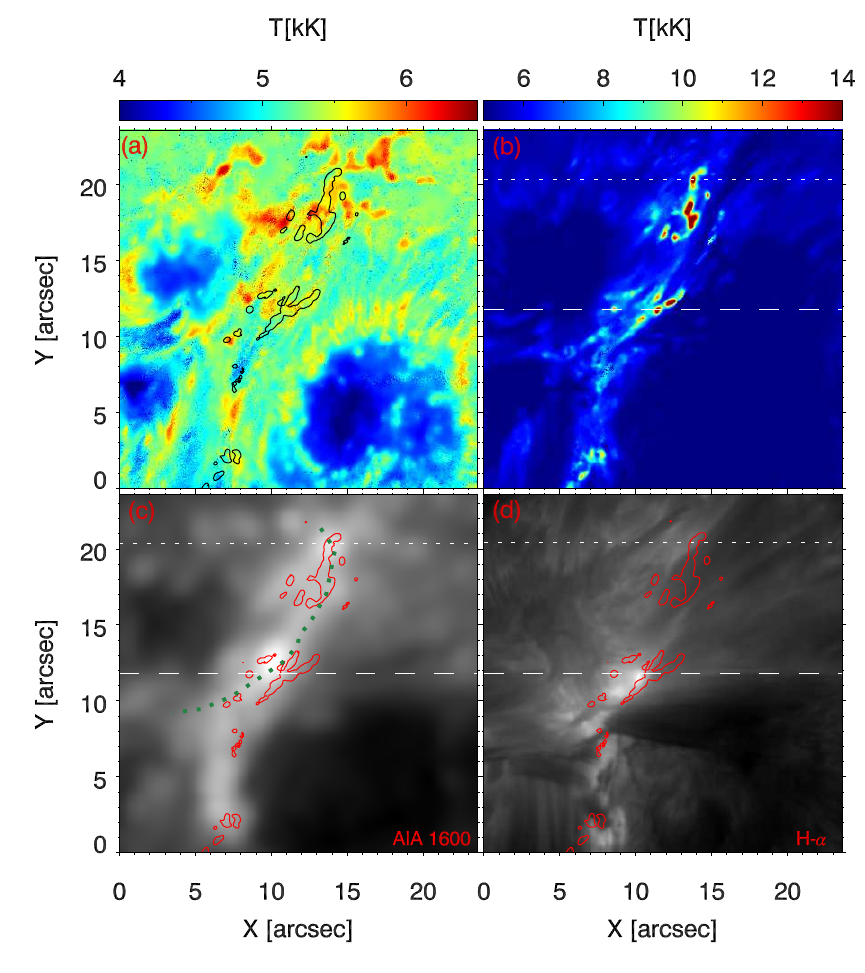}\caption{ Atmospheric temperature and AIA 1600~$\AA$ and \halpha\ brightness in the emerging loop. 
(a): photospheric temperature map at $\log \tau_{500} = -1$  from the \FeI\ 6301-6302~\AA\ inversion. 
(b): chromospheric temperature map at $\log \tau_{500} = -3$ from the \CaII ~8542~\AA\ inversion.
(c): AIA 1600~$\AA$ intensity on a logarithmic brightness scale.
(d): \halpha\ line-core intensity.
The irregular coutours enclose locations with $T>8$~kK in the $\log \tau_{500} = -3$ map.
The dotted and dashed horizontal lines are the cuts shown in Figure~\ref{fig:temp_p_cut_111_114}.
The green dotted line is the path shown in Figure~\ref{fig:velblob}}\label{fig:tempvssdo_111_114} 
\end{figure}

\begin{figure}
\includegraphics[scale=1]{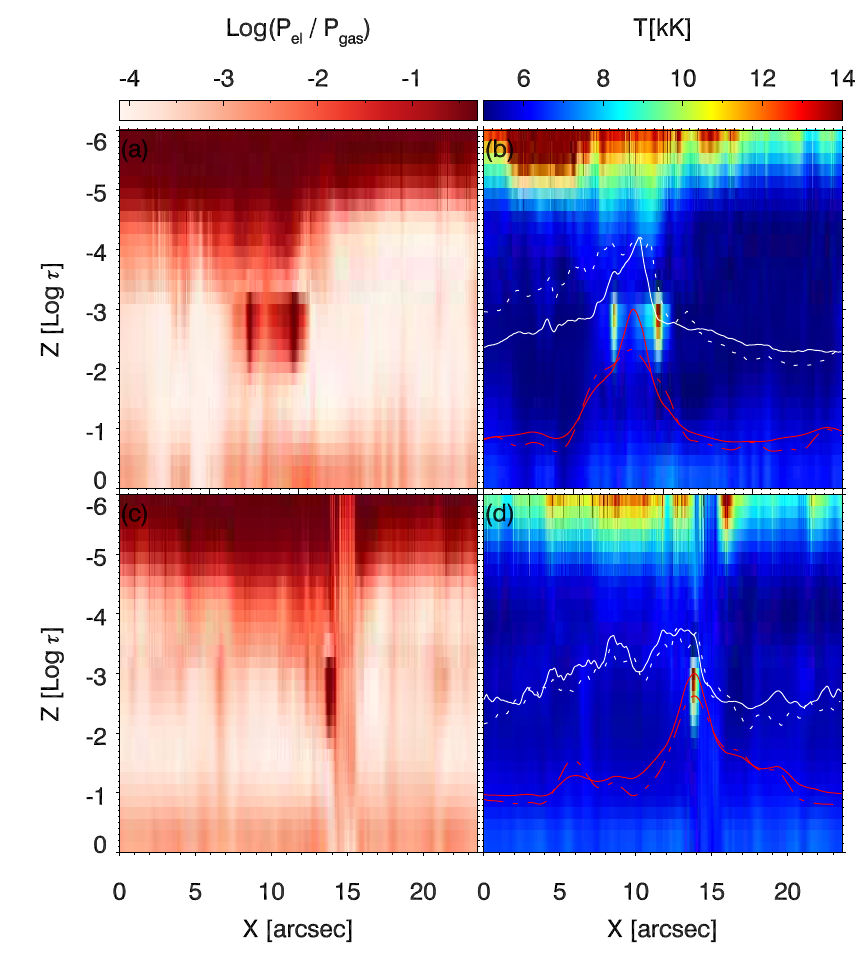}\caption{\textit{Left-hand column}: vertical cut of the ratio between the electron and gas pressure at $Y=12\arcsec$ (a) and at $Y=20\arcsec$ (c) 
\textit{Right-hand column}: vertical cut of the temperature at  $Y=12\arcsec$ (b) and at $Y=20\arcsec$ (d). 
The white solid curve is the \halpha\ line-core intensity, the white dotted line the CaII ~8542~\AA\ core intensity, the red solid line the intensity curve for AIA~1600~\AA\ and the red dotted line the AIA~1700~\AA\ intensity. The locations of the cuts are indicated in Figure~\ref{fig:tempvssdo_111_114}. } \label{fig:temp_p_cut_111_114} 
\end{figure}

Figure~\ref{fig:tempvssdo_111_114} shows the temperature maps retrieved by the inversion. The photospheric map (a) at $\log \tau_{500} = -1$ and the chromospheric one (b) at $\log \tau_{500} = -3$ are obtained from \FeI\ ~6301-6302~\AA\ and \CaII ~8542~\AA\ respectively. The contours in panels a, c and d indicate the region of the chromospheric map where the temperature exceeds 8~kK at $\log \tau_{500} = -3$. The photospheric temperature map shows neither sign of strong heating nor of correlation with the temperature in the chromosphere. This is quite expected since no intensity enhancement is observed in Figure~\ref{fig:overview}-a. 
In the chromosphere, the penumbra  harbours hot patches at the location of the jet footpoints, as already shown in Figure~\ref{fig:magnetograms_111_114}-i. According to the  \CaII ~8542~\AA\  inversion results, these regions have temperatures up to 14 kK. A similar temperature pattern can be recognised in the  co-aligned AIA~1600  intensity map (c). Two round brightenings at $Y=12\arcsec$ and $Y=20\arcsec$  coincide with the highest temperature patches in the chromospheric map. However, the bulk of the brightening of both AIA~1600 (c) and \halpha\ (d), at $Y=12\arcsec$, is not exactly cospatial with the hot patches retrieved from the inversion, but is instead located in between the patches.

The right-hand column of Figure~\ref{fig:temp_p_cut_111_114} shows the temperature cuts along $Y=12\arcsec$ and $Y=20\arcsec$ (long and short-dashed lines respectively in Figure~\ref{fig:tempvssdo_111_114}). The increased temperature is very concentrated in height around  $\log \tau_{500} = -3$ has an horizontal extent less than 1$\arcsec$. The vertical cut at $Y=20\arcsec$  shows similar behaviour in terms of height and size. In Figure~\ref{fig:temp_p_cut_111_114}, we have plotted along the two cuts the intensity of \halpha\ line core (white solid line), \CaII ~8542~\AA\ line core (white dotted line), AIA~1600 (red solid line) and AIA~1700 (red dashed line). The four intensity curves are all normalised by their own maximum value and multiplied by a common scale factor to match the figure size. 

At $Y=12\arcsec$ (b), the location of the heating corresponds to an increase of the \halpha\ intensity. The highest \halpha\ intensity is however located between the two heated regions, where the temperature estimated from \CaII ~8542~\AA\ is almost 5~kK less. Also the AIA~1600 intensity curve is peaked at the same location. AIA~1700 shows there a bump that is however less pronounced. This suggests that the \CaII ~8542~\AA\ line could be blind to the possibly very high temperature between the two heating regions because \CaII\ has been ionized away. Another hint for a hidden heating is that, as shown in Figure~\ref{fig:shear}, the direction of the magnetic field has a significant vertical gradient in the region between the two patches (between $X=8\arcsec$ and $X=12\arcsec$). One would expect therefore to see the sign of a possible field shear  but this is missing. 

The region corresponding to the Y=12$\arcsec$ cut appears in \CaII ~8542~\AA\ as diffusely bright, and we do not observe a jump in its intensity curve as strong as in \halpha\, on the left side.
The abrupt drop  of the \halpha\ and \CaII ~8542~\AA\ intensity curve on the right side is due to the presence of the opaque jets. This drop is  shifted by $\sim1\arcsec$ with respect to the hot region. A projection effect, due to the difference in height between \CaII ~8542~\AA\ and \halpha, can however be excluded, because \CaII ~8542~\AA\ does not look aligned with the heating.

The line core intensities of \halpha\ in panel d have a spatially larger peak than the width of the heating region and its maximum value is 80\% of the maximum along the Y=12$\arcsec$ cut. Similar behaviour is shown also by \CaII ~8542~\AA. Most of the area subtended by this peak has a low temperature (6-7~kK). The heating extent is better matched by the AIA~1600 curve and and also by AIA~1700, which, as in panel b, has a less intense jump than AIA~1600.

The left column of Figure~\ref{fig:temp_p_cut_111_114} shows the logarithm of the ratio between the electron and the gas pressure for the two vertical cuts so far considered. Here we see that, at the same location where heating takes place, the electron pressure grows up to values comparable with the gas pressure, showing that the plasma is strongly ionized there. The biggest contribution to the electron pressure is given by hydrogen. 

\begin{figure}
\includegraphics[scale=1]{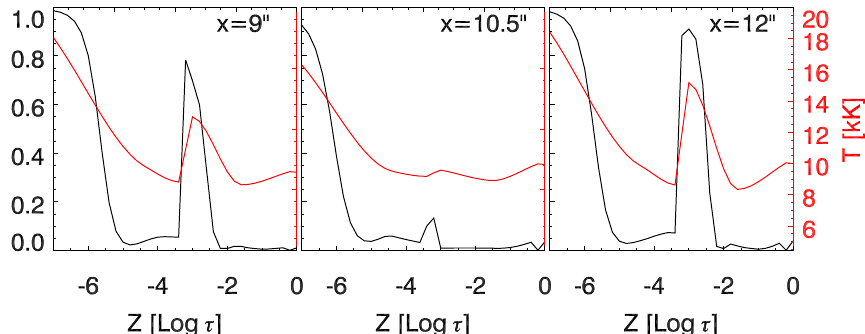}\caption{Temperature (red) and the fraction of calcium as \CaIII\ (black) at three locations along the $Y=12 \arcsec$ cut of Fig.~\ref{fig:temp_p_cut_111_114}. Left: first heating region at $X=9\arcsec$. Middle: the low temperature region between the two heating regions at $X=10.5\arcsec$. Right: second heating region at $X=12\arcsec$.} \label{fig:popul} 
\end{figure}

To evaluate the ionisation of \CaII\, we have computed the populations for each level of a 5-level atom of \CaII\ plus the \CaIII\ continuum using NICOLE. The fraction of all calcium in the form of \CaIII, as well as the temperature are given as a function of height in Figure~\ref{fig:popul} at three different pixels along the Y=12$\arcsec$ cut. The first and the last panels coincide with the heating locations (X=9$\arcsec$ and X=12$\arcsec$). The locations with strong heating are located around $\log \tau_{500} = -3$. As expected, \CaII\ is largely ionized away at these locations.

At the point between the two heating locations (X=10.5$\arcsec$), the inversion predicts a low temperature and a lower ionization of \CaII\ is less ionised. As suggested by the strong 1600~\AA\ emission compared the the 1700~\AA\ emission we interpret this low-temperature area as an artefact of the \CaII-based inversion, which is insensitive to temperatures well above 15~kK because the \CaII\ is ionized away. In reality the temperature will be substantially higher than 15~kK in order to produce the 1600~\AA\ emission.

%

\subsubsection{Velocity}

\begin{figure}
\includegraphics[scale=1]{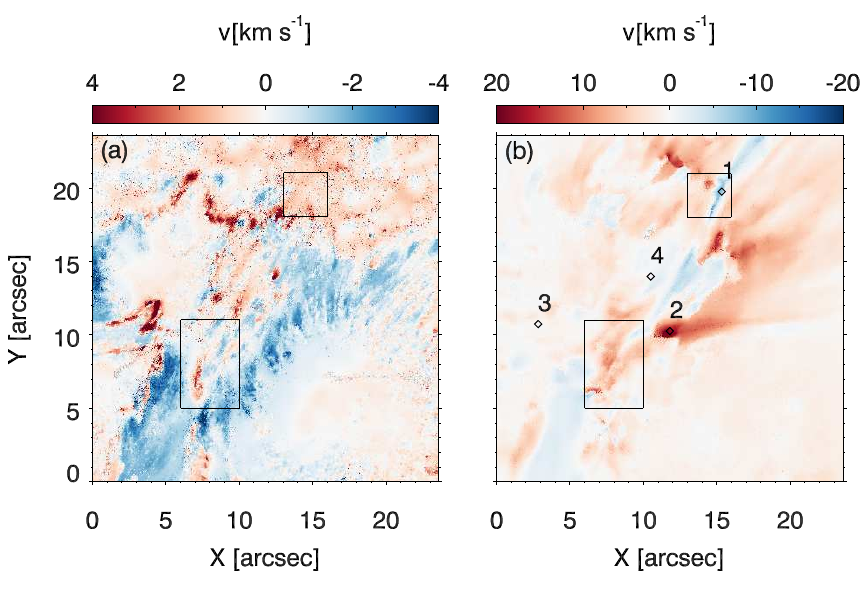}\caption{Velocity map of (a) the photopshere at $\log \tau_{500} = -1$, (b) chromosphere at $\log \tau_{500} = -3$. The numbered diamonds indicate the location of the selected Stokes profiles of Figure~\ref{fig:stokes}. The two rectangles show the locations of the footpoints of the emerging loop structure.}\label{fig:vel_111_114}
\end{figure}

Figure~\ref{fig:vel_111_114} shows the LOS velocity maps calibrated using as reference the velocity in the umbra  \citep[see][]{1977ApJ...213..900B}.
As expected from the intensity in Figure~\ref{fig:overview}-a, the photosphere (Figure~\ref{fig:vel_111_114}-a) does not show any velocity pattern connected to the presence of the jet. The rectangles highlights the position of the loop-structure footpoints. In one of the footpoint of the loop  structure (bottom), where the inclination is close to -80$^\circ$, there is a downward LOS velocity of 4~\kms, that can be connected to the plasma drain along the loop structure.

The upwards and downwards motion of the jets is retrieved in the chromospheric velocity map (Figure~\ref{fig:vel_111_114}-b). The jets have LOS velocities of the order of 10-20 \kms. An inspection of the time-length diagram, performed with the CRisp SPectral EXplorer
\citep[CRISPEX,][]{2012ApJ...750...22V} along different jet paths close to the one indicated in Figure~\ref{fig:magnetograms_111_114}, shows that the jets have a plane-of-the-sky (POS) velocity between 100-200~\kms. Due to the low time cadence (27~s), an estimate of the LOS velocity, with the same approach, is unfeasible. 
Comparing the LOS velocity obtained by inversion with the POS velocity, we notice that the POS component dominates. This suggests that the jets on the investigated paths may actually follow a rather horizontal trajectory.

\begin{figure}
\includegraphics[scale=1]{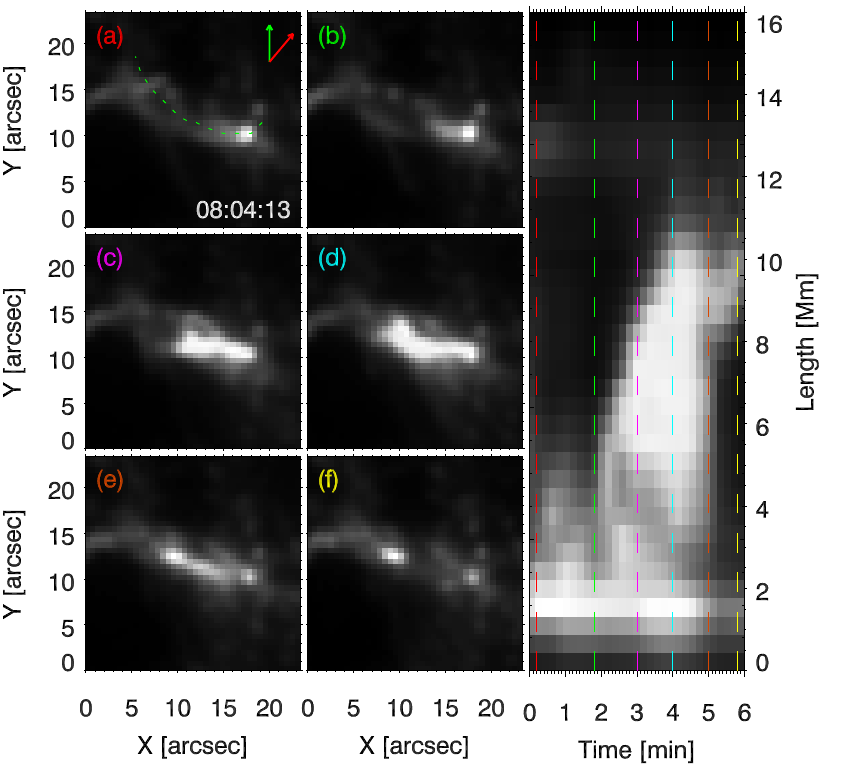}\caption{Intensity maps and a space-time diagram of the jet footpoints in the AIA~1600 channel. Panels (a)--(f) show the maps at equidistant time intervals between 08:04:13~UT and 08:09:49~UT. The right column shows the space time diagram along the path traced by the green dotted line in panel (a). The times of the maps are indicated in the space-time digram with dashed colored lines. The green and the red arrows point towards the \grass{north} and the disk centre respectively.}\label{fig:velblob}
\end{figure}

From the on-line animation of Figure~\ref{fig:overview}, it is clear that the jets not only have a vertical motion but also a sideways motion. It is possible to observe a bright blob moving along the footpoint line (E-W direction). To estimate the transversal motion on the POS, we have selected 30 frames of AIA~1600. 
Figure~\ref{fig:velblob}-a-f displays six of these frames. 
In the non-spatially-aligned AIA~1600 dataset, although having a lower spatial resolution than CRISP, the high-temperature moving blob can be more easily identified. The continuously tenuously  bright structure in panels a-f is the jet footpoint line. Superimposed on it we observe an enhancement of intensity expanding on both sides of a ribbon-like structure, until it faints (e) and only two end points are left visible (f). This ribbon-like structure appears in the AIA~1600 channel but not in \halpha.  A faint trace of the jets is visible in AIA~1600, especially in the animation of Figure~\ref{fig:allchannels}, and it moves coherently with the brightening, apparently at the same speed.
The rightmost panel of Figure~\ref{fig:velblob} shows the space-time diagram traced along the green path of panel a. This path aims to follow the motion of the blob and it has also been plotted in the CRISP-aligned image of Figure~\ref{fig:tempvssdo_111_114}-c to easily frame its location.
From the time-space diagram of the six different time steps,  we can read that the blob - and with it the jets- moves along the footpoint line at almost 45~km~s$^{-1}$. This value will be useful in the comparison with other types of transversal motion and their production mechanisms.

\subsection{Time evolution}
\begin{figure}
\includegraphics[scale=1]{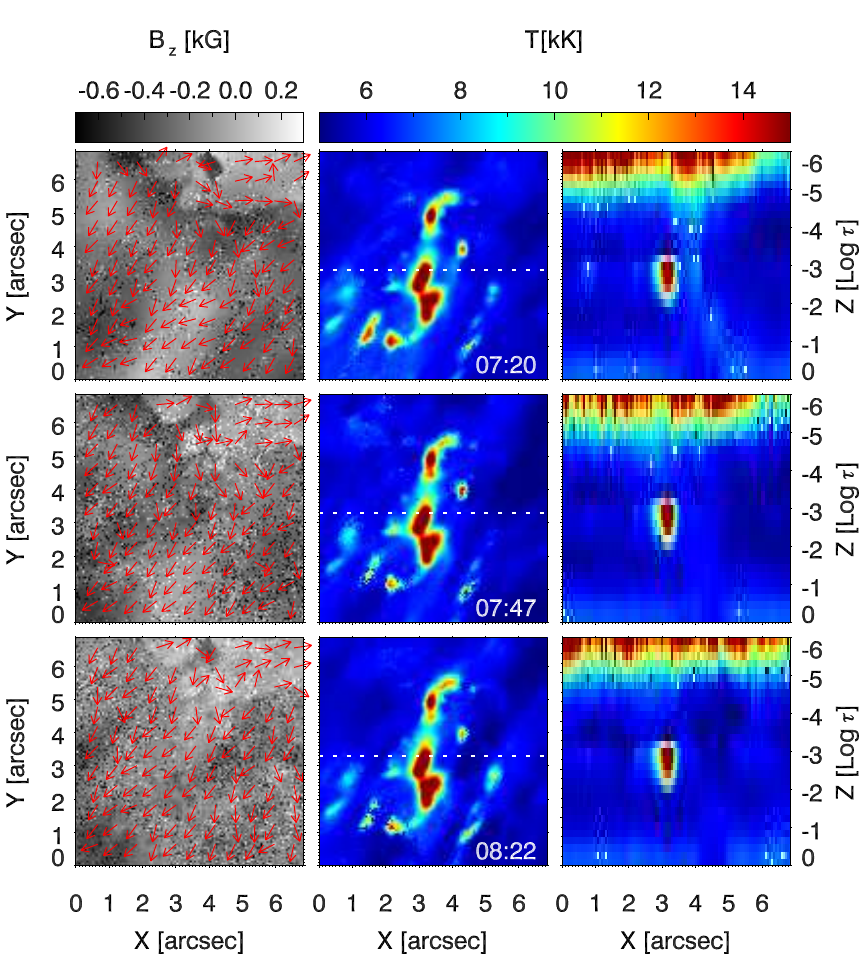}\caption{Time evolution of the upper small field of view (see Figure~\ref{fig:overview}-d). \textit{Left column:} vertical magnetic field map and the red arrows are the direction of the horizontal field. \textit{Middle column:} temperature map at $\log \tau_{500} = -3$. \textit{Right column:} vertical cut of the temperature, same cut of Figure~\ref{fig:temp_p_cut_111_114}. Every row corresponds to a different time step, \textit{the first row} is the beginning of the observations, \textit{the second row} the middle of the time series and \textit{the last row} the end of the observations.}\label{fig:up_timeevolution}
\end{figure}

\begin{figure}
\includegraphics[scale=1]{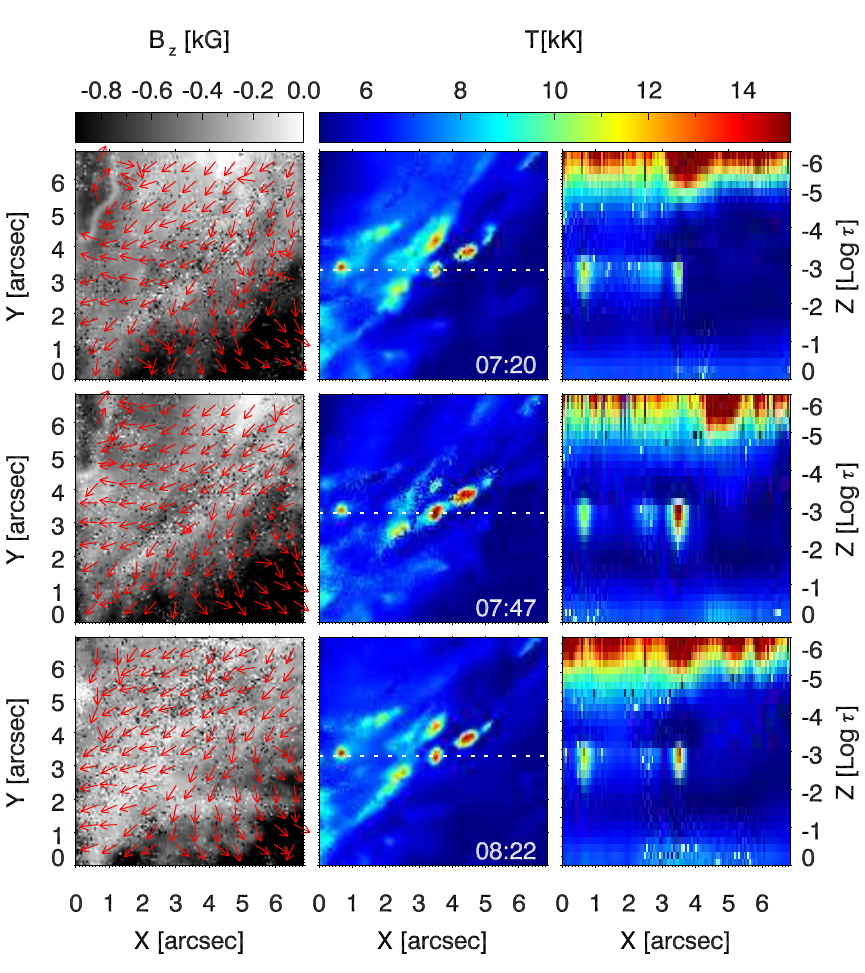}\caption{Same as Figure~\ref{fig:up_timeevolution}, but now showing the time evolution of the lower small field of view (see Figure~\ref{fig:overview}-d).}\label{fig:bottom_timeevolution}
\end{figure}

The smaller FOVs, selected for the complete time series inversion (Figure~\ref{fig:overview}-d), contain the location where the magnetic reconnection is supposed to take place. Unfortunately our observations last only for 66~minutes and they do not capture either the beginning nor the end of the jets, which last for almost 4 hours. Therefore the picture that is retrieved by the time evolution is rather stable. The magnetic field does not show any clear periodicity and it has a quite constant value and direction. This stability is exemplified by Figure~\ref{fig:up_timeevolution} and \ref{fig:bottom_timeevolution}, which show the maps of some key quantities retrieved by the inversion of different time steps. We have decided to show the vertical magnetic field and the direction of the horizontal field at beginning (first row), in the middle (middle row) and at the end of the time series (last row). The only clear change in the $B_z$ map is due to the motion of the jets which appears in  Figure~\ref{fig:bottom_timeevolution} as different branches of weak field between X=3 and X=6. The temperature in both the FOVs shows a constant pattern. As in the case of $B_z$, variation of temperature can be found in the lower FOV (Figure~\ref{fig:bottom_timeevolution}) close to the hotter regions due to the motion of the cold jets. In general, the location of the heating is constant throughout the entire time series between $\log \tau_{500} = -2$ and $\log \tau_{500} = -3$. In the XY plane the heating is confined in the same region outlined by the contour in Figure~\ref{fig:magnetograms_111_114}. 

The stable configuration that arises from inversion is in agreement with observations of a quasi-stationary phenomenon, but it is also  surprising when compared with the fast motion of the brightenings at the jet footpoints (see for instance Figure~\ref{fig:velblob}). This may suggest a different origin or mechanism for the brightenings steadily located at the heating locations and those moving along the jet footpoints.

\section{Discussion \& Conclusions}

We have studied the atmosphere above a $\delta$-sunspot that harbours a series of recurrent fan-shaped jets in its penumbra. We applied  non-LTE inversion techniques to the spectropolarimetric data to retrieve chromospheric quantities and their photospheric counterparts. 

The photospheric magnetic field map shows a secondary pair of polarities located in the penumbra shared by the main negative umbra and the two smaller spots.
The scenario that we obtain is in agreement with an emerging magnetic loop structure in the penumbra of the sunspot. The footpoints of this structure are well visible in the photosphere (Figure~\ref{fig:magnetograms_111_114}-f) while the chromosphere shows the most horizontal part. The presence of a secondary bipolar structure has been simulated by \citet{2015ApJ...813..112T} but the configuration that we have obtained is hard to reconcile with a single buoyant structure. Moreover, a downflow is expected between the second polarities but the photospheric velocity panel of Figure~\ref{fig:vel_111_114} proves it wrong. 

Magnetic reconnection between the emerging loop and the surrounding pre-existing penumbral field is likely to take place and produce the brightenings observed at the jet footpoints. 
To this location corresponds an increase of the temperature up to 14~kK at $\log \tau_{500} = -3$  and a strong increase of the electron pressure. Much of the plasma is ionised and visible also in AIA 1600~\AA. 
We therefore conclude that the plasma temperature at certain locations in the chromosphere is higher than the value retrieved by the \CaII\ inversion. The heating is localised between $\log \tau_{500} = -2$ and 
$\log \tau_{500} = -3$, that is rather deep in the chromosphere. 

This result poses a new boundary on the height at which the magnetic reconnection takes place. Therefore, it would be interesting to reproduce these fan-shaped jets via MHD simulations, starting from the magnetic configuration that has been show in this paper. This would imply the usage of a strong (>1kG) and vertical field as guide field and the emergence of a horizontal weaker field. Like in previous observations with similar topology \citep{2016A&A...590A..57R}, the field that we have obtained here does not show sign of twist. 
So, an obvious choice would be to recreate at the photosphere an untwisted or slightly twisted magnetic field, as simulated by \citet{2015ApJ...811..138T} for their light-bridge configuration.

So far no model or simulation has been able to recreate the exact magnetic topology and the dynamics of these fan-shaped jets.
In particular, the jets exhibit a sideways motion that is absent in the simulation of \citet{2011ApJ...726L..16J}. The values that we have obtained here are in the order of magnitude of those found for anemone jets: $10-20$~\kms\ from \citet{1992PASJ...44L.173S}, $0-35$~\kms\ from \citet{2007PASJ...59S.771S} and $\sim 10$~\kms\ from \citet{2008ApJ...673L.211M}. This suggests that the mechanism behind this may be similar and therefore that the sideways motion in the fan-shaped jets of this paper could be explained by the emergence of the loop structure and its reconnection. This mechanism has been shown by \citet{2008ApJ...673L.211M} for the emergence of a dome structure.


While the magnetic field configuration at the jet footpoint appears clear, the jet bulk does not have a polarisation signal.
We have observed the imprint of the jet in some of those that are dark in \CaII ~8542~\AA. The inclination that we have estimated from the magnetic field vector is rather horizontal. As we have already mentioned, the field value at this location is not reliable and consequently the direction of the jet too. However, from the general context  we can see that an horizontal inclination can be reconciled with the fact that the jet trajectories lie indeed on the direction of the coronal loop connecting the main polarities of the active region (see Figure~\ref{fig:loop}). Since the footpoints of the jets are located relatively low in the atmosphere compared to the top of the loop it is reasonable to think that at chromospheric height the loop is rather stretched and horizontal.

\begin{acknowledgements}
The Swedish 1-m Solar Telescope is operated by the Institute for Solar
Physics of Stockholm University in the Spanish Observatorio del Roque
de los Muchachos of the Instituto de Astrof\'{\i}sica de Canarias. 
The computations were performed on resources provided by the Swedish National Infrastructure for Computing (SNIC) at the High Performance Computing Center North at Ume{\aa} University.
JdlCR is supported by grants from the Swedish Research Council (2015-03994), the Swedish National Space Board (128/15) and the Swedish Civil Contigencies Agency (MSB). This research was supported by the CHROMOBS and CHROMATIC grants of the Knut och Alice Wallenberg foundation.
\end{acknowledgements}


\bibliographystyle{aa}

\end{document}